\documentclass[twocolumn,amsmath,amssymb,prl,superscriptaddress,longbibliography]{revtex4-1}

\def\vec{\mathbf}

\usepackage{titlesec}
\usepackage[citecolor=blue,colorlinks=true,linkcolor=blue]{hyperref}
\usepackage{epsfig}
\usepackage{color}
\usepackage{graphicx} 
\usepackage{dcolumn}  

\newcommand{\be}{\begin{equation}}
\newcommand{\ee}{\end{equation}}

\begin{document}

\title{Pressure-tuned magnetic interactions in honeycomb Kitaev materials}

\author{Ravi Yadav}
\affiliation
{Institute for Theoretical Solid State Physics, IFW Dresden, Helmholtzstr. 20, 01069 Dresden, Germany}

\author{Stephan Rachel}
\affiliation{School of Physics, University of Melbourne, Parkville, VIC 3010, Australia}

\author{Liviu Hozoi}
\affiliation
{Institute for Theoretical Solid State Physics, IFW Dresden, Helmholtzstr. 20, 01069 Dresden, Germany}

\author{Jeroen van den Brink}
\affiliation
{Institute for Theoretical Solid State Physics, IFW Dresden, Helmholtzstr. 20, 01069 Dresden, Germany}
\affiliation{Department of Physics, Technical University Dresden, 01062 Dresden, Germany}

\author{George Jackeli}
\altaffiliation[]{Also at Andronikashvili Institute of Physics, 0177
Tbilisi, Georgia}
\affiliation{Institute for Functional Matter and Quantum Technologies, 
University of Stuttgart, Pfaffenwaldring 57, D-70569 Stuttgart, Germany}
\affiliation{Max Planck Institute for Solid State Research,
Heisenbergstrasse 1, D-70569 Stuttgart, Germany}

\begin{abstract}
%
A range of honeycomb-lattice compounds has been proposed and investigated in the search for 
a topological Kitaev spin liquid. 
However, sizable Heisenberg interactions and additional symmetry-allowed exchange anisotropies in the magnetic Hamiltonian of these potential Kitaev materials push them away from the pure Kitaev spin-liquid state. Particularly the Kitaev-to-Heisenberg coupling ratio is essential in this respect.
With the help of advanced quantum-chemistry methods, we explore how the magnetic coupling ratios depend on pressure in several honeycomb compounds (Na$_2$IrO$_3$, $\beta$-Li$_2$IrO$_3$, and $\alpha$-RuCl$_3$).
We find that the Heisenberg and Kitaev terms are affected differently by uniform pressure or strain: the Kitaev component increases more rapidly than the Heisenberg counterpart. This provides a scenario where applying pressure or strain can stabilize a spin liquid in such materials. 
\end{abstract}

\maketitle

{\it Introduction.\, }
The realization of quantum spin liquids (QSLs) in spin-orbit driven correlated materials is an intensively pursued goal in the condensed matter community, both experimentally and theoretically.
In a QSL strong quantum fluctuations  prevent long-range magnetic order even at the lowest temperatures and instead a non-trivial ground state forms with long-range quantum entanglement between spins\,\cite{balents10,han12,yamashita10}.
Of particularly great promise in this context is the Kitaev Hamiltonian on honeycomb lattices, which exhibits various topological spin-liquid phases\,\cite{Kitaev06}. 
The paramount attention given to such states can be understood by the fact that they are topologically protected from decoherence\,\cite{Albrecht16}, display fractional excitations with
Majorana statistics, and therefore hold promise in the field of quantum information and quantum computation.

The quest for the physical realization of the Kitaev spin liquid for effectively spin-1/2 particles took a big stride forward with the proposal of the honeycomb 5$d^5$ iridate materials as host of the Kitaev-Heisenberg model\,\cite{Jackeli09,Chaloupka10}. 
The latter describes the interactions between spin-1/2 moments with the help of two competing nearest-neighbor (NN) couplings, i.e., an isotropic Heisenberg term ($J$) assumed to mainly arise from direct exchange between Ir-ion $d$ orbitals and an anisotropic Kitaev component ($K$) which stems from superexchange along the Ir-O-Ir paths.

Certain materials, in particular Na$_2$IrO$_3$, $\beta$-Li$_2$IrO$_3$, and $\alpha$-RuCl$_3$, have been extensively studied experimentally in this context\,\cite{Singh10,Ye12,Choi12,Modic14,Takayama15,Chun15,johnson15,sandilands15,Sandilands16,banerjee16,Banerjee17} 
as well as  within the electronic-structure computational field, by either quantum-chemistry~\cite{Katukuri14,Katukuri16,Yadav16} or density-functional-based~\cite{rau14,Yamaji14(9),Kim2015,Winter2017} methods. However, it turns out that the anticipated spin-liquid regime is precluded in these honeycomb compounds, most likely
 due to 
 the presence of reasonably strong Heisenberg interactions, longer-range spin couplings, or the combination of both these factors.  
So far, all the measurements indicate magnetic long-range order at low temperatures and zero external magnetic field. None of these systems however exhibits 
the conventional N\'eel state although the magnetic ions form bipartite lattices in all of them. It has been suggested that these materials are still located in the phase diagram in close vicinity to the spin-liquid regime\,\cite{Takayama15,Chun15,sandilands15,banerjee16,Banerjee17}.
This has then inspired rigorous experimental effort to test their properties under strain or pressure~\cite{Takayama15,Veiga17,Nicholas17,Wang17}.
In particular, there have been claims for finding the evidences of spin-liquid states under applied pressure in 
$\beta$-Li$_2$IrO$_3$~\cite{Takayama15,Veiga17}, $\gamma$-Li$_2$IrO$_3$~\cite{Nicholas17}, and $\alpha$-RuCl$_3$~\cite{Wang17}.
It is worth noting that even more complex strain experiments have been suggested\,\cite{rachel16,perreault17}.
For $\alpha$-RuCl$_3$ indications for an emergent spin-liquid phase induced by magnetic field have also been observed~\cite{Baek17,Yadav16}.

Here we explore the effects of pressure on the NN isotropic and anisotropic interactions by employing {\it ab initio} quantum-chemistry methods. 
In order to interpret the response of the magnetic exchange couplings under uniform pressure, we start by analysing the dependence of both the Kitaev and Heisenberg terms on the lattice constants.
By adopting an idealised model in which under uniform pressure the lattice just scales down to a configuration with smaller unit-cell parameters, we obtain expressions which show that the Kitaev coupling constant increases
more rapidly than the Heisenberg $J$, giving large $K/J$ ratios for shorter bonds and thus an enhancement towards spin-liquid formation in the phase diagram of the Kitaev-Heisenberg model under volume change. 
The amplitude of this enhancement for the Kitaev term is also confirmed by {\it ab initio} quantum-chemistry calculations in the case of honeycomb compounds. However, in the case of 
hyperhoneycomb Li$_2$IrO$_3$, in addition to the upsurge of Kitaev exchange, the symmetric off-diagonal $\Gamma$ couplings also become significantly larger and might play an important role in shaping its magnetic properties, as discussed in Ref.\,\onlinecite{Ioannis17}.
Looking at such trends gives a profound insight into the different competing processes coming into play for different compounds or structures and can provide guidelines or direction for further experimental investigations.


{\it Qualitative analysis.\, }
 The Kitaev-Heisenberg Hamiltonian~\cite{Chaloupka10} originally proposed as a minimal model for the honeycomb-lattice iridates takes the following form on a given bond of NN's $i$, $j$:
\begin{equation}
 {\cal H}^{(\gamma)}_{ij} =J\, \tilde{\bf{S}}_i \cdot \tilde{\bf{S}}_j
           +K \tilde{S}^\gamma_i \tilde{S}^\gamma_j,
\label{Eq:ham0}
\end{equation} 
where $\tilde{\bf{S}}_i$ and $\tilde{\bf{S}}_j$  represent pseudospin 1/2 operators for the ground-state Kramers doublets 
of Ir$^{4+}$ (or Ru$^{3+}$) ions, the first and second terms correspond to  the isotropic Heisenberg interaction and the anisotropic Kitaev coupling, respectively, and $\gamma$ $\in$ $\{x,y,z\}$ labels the three
inequivalent bonds and the corresponding Cartesian components of the pseudospins.
Depending on the $K/J$ ratio, the model (\ref{Eq:ham0})  is known to host a rich phase diagram containing the Kitaev spin-liquid and  a variety of ordered states~\cite{Chaloupka10,Chaloupka13}.  

To qualitatively understand pressure effects on the effective coupling constants $K$ and $J$, we assume that under uniform pressure all inter-atomic distances rescale in the same way and
consider only the leading contributions to the exchange interactions.
A perturbative analysis estimates that $J \sim \frac{t_{dd}^2}{U}$ and $K\sim -\frac{t_{pd}^4}{\Delta_{pd}^2} \frac{J_H}{U^2}$~\cite{Jackeli09,Chaloupka10}, with the Heisenberg term
predominantly related to direct exchange while the Kitaev interaction is mostly due to superexchange processes along the Ir-O-Ir paths.
Here, $t_{dd}$ and $t_{pd}$ stand for the hybridisation amplitudes between $d$-orbitals  of neighboring Ir ions and between Ir $d$ and O $p$ states,
respectively, and $\Delta_{pd}$ is the charge-transfer energy. The interaction parameters $U$ and $J_H$ correspond  to the on-site Coulomb repulsion and the
Hund coupling, respectively. In the simplest picture, the hybridisation amplitudes scale with the inter-ionic distance $r$ as $t_{dd}\sim r^{-5}$ and $t_{pd}\sim r^{-7/2}$~\cite{Har89}.
This, in turn, gives a rescaling of the coupling constants $J= {J_0}\left(\frac{a}{a_0}\right)^{-10}$ and $K={K_0} \left(\frac{a}{a_0}\right)^{-14}$ when the characteristic inter-ionic length
scale changes from $a_0$ to $a$ under uniform pressure or strain.

%
%
%
%
%
The above naive estimates are based on the dominant subset of possible exchange processes and are thus rather rough in character.
However, they do suggest that the strengths of the different NN isotropic and anisotropic coupling constants  get
differently renormalized under uniform pressure.
In order to test this quantitatively we have performed electronic-structure calculations using many-body quantum-chemistry methods to rigorously account for 
all symmetry-allowed NN exchange paths for variable inter-ionic distances within a family of potential Kitaev spin-liquid materials.

 \begin{figure}
 \begin{center}
 \includegraphics[width=7.4cm,keepaspectratio=true]{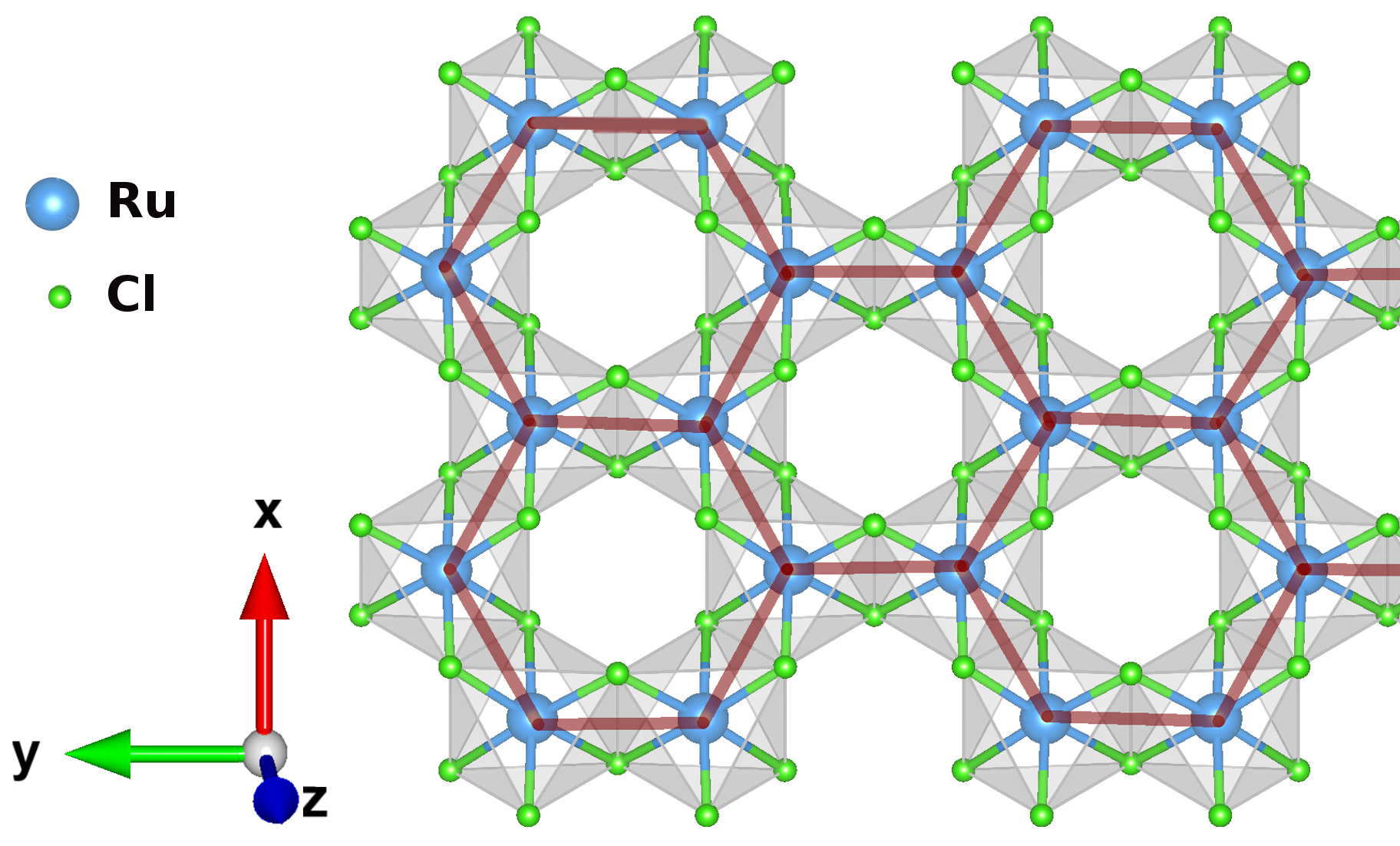}\\[10pt]
%
 \includegraphics[width=7.4cm,keepaspectratio=true]{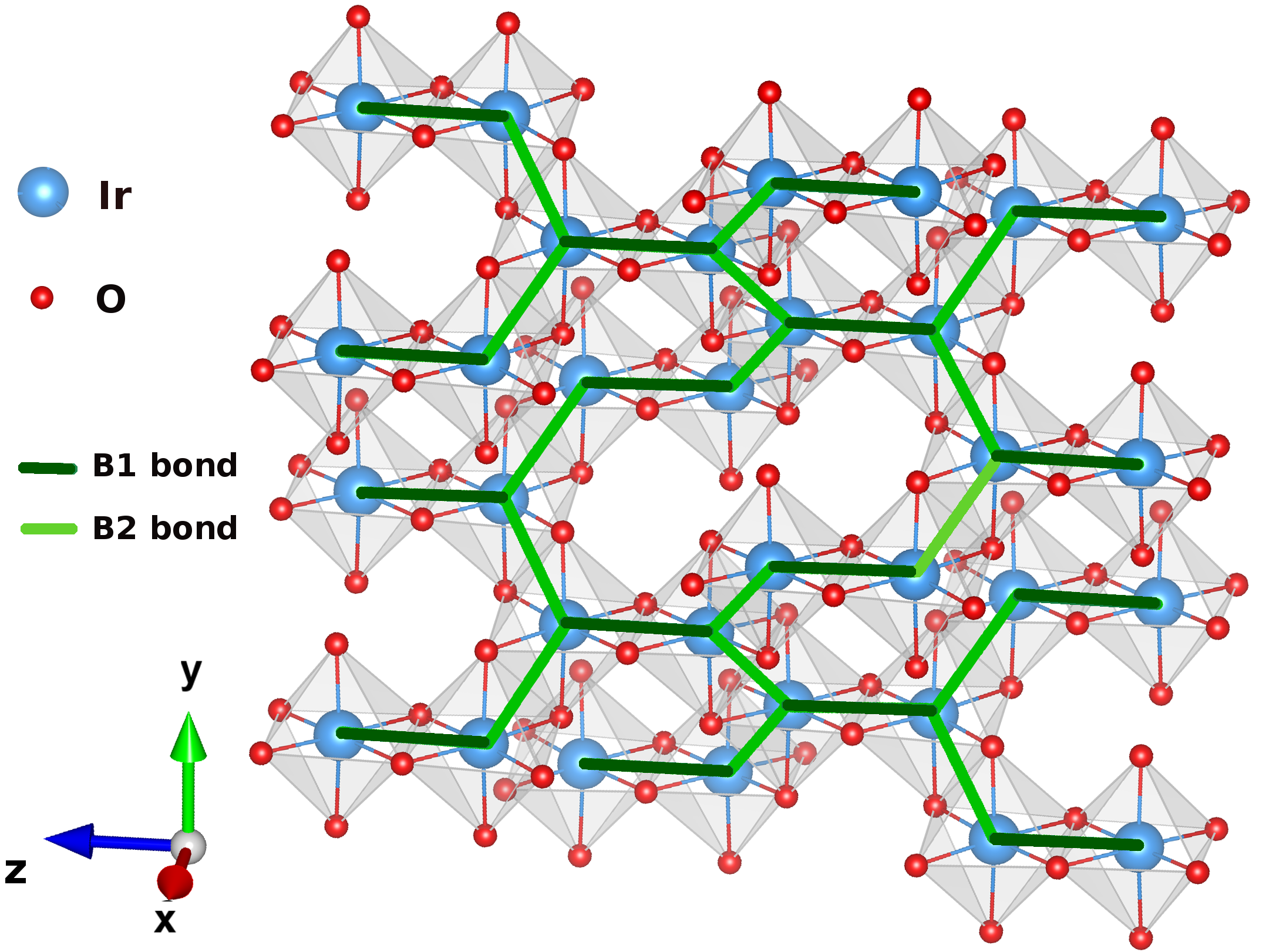}
\end{center}
 \caption{(top) Ru-ion honeycomb lattice (blue) with Cl-ligand octahedral coordination (green sites) in RuCl$_3$. (bottom) Ir-ion hyperhoneycomb lattice in $\beta$-Li$_2$IrO$_3$. The local environment
of the Ir sites remains similar to the 2D honeycomb network shown above.}
\end{figure}

{\it Electronic-structure calculations.}
The transition-metal (TM) ions, i.e., Ir/Ru, in the oxides and the chloride discussed here frame a honeycomb (Na$_2$IrO$_3$ and $\alpha$-RuCl$_3$) or hyperhoneycomb lattice ($\beta$-Li$_2$IrO$_3$), as shown in Fig.\,1, where each site is connected to three TM NNs. 
The Ir ions in Na$_2$IrO$_3$ and $\beta$-Li$_2$IrO$_3$ have a 4$+$ oxidation state, which implies five 5$d$-shell electrons. 
The octahedral ligand (L) coordination makes that 
the $5d$ levels are split into $e_g$ and $t_{2g}$ states, with the latter lying at significantly lower energy~\cite{Gretarsson13}. 
Given the large $t_{2g}$--$e_{g}$ splitting, the five valence electrons enter the $t_{2g}$ subshell in first approximation, which yields an effective picture of one hole within the $t_{2g}$ sector. 
In the presence of strong spin-orbit coupling, this can be mapped onto a set of fully occupied $j_{\rm eff} =$ $3/2$ and half-filled $j_{\rm eff} =$ $1/2$ states~\cite{Jackeli09, Abragam70, Kim08}.
Deviations from a perfect cubic environment may lead to admixture of these $j_{\rm eff}=1/2$ and $j_{\rm eff}=3/2$ components. 
The key structural difference between the honeycomb and hyperhoneycomb lattices is that the Ir ions frame a truly 2D network in the former while
they form a slightly more complicated 3D arrangement in the latter, with alternate rotation of two adjacent B2 bonds around the B1 link (see Fig 1(b)). However, the essential local environment of Ir is similar to the 2D honeycomb structure. 
The Ru ions in $\alpha$-RuCl$_3$ have a 3$+$ oxidation state, which again implies five electrons in the ($4d$) valence shell.
Similar to the 5$d$ compounds, the ligand field splits the $4d$ levels into $t_{2g}$ and $e_{g}$ states.
Spin-orbit coupling is significantly weaker for $4d$ orbitals, but still large enough to generate strongly spin-orbital entangled 1/2 pseudospins for moderate
trigonal distortion\,\cite{Yadav16}.

\begin{table}[b]
\caption{NN magnetic couplings (in meV) for bond B1 in Na$_2$IrO$_3$ as functions of variable Ir-Ir bond length $a$; the relative change is $\delta a = a/a_0-1$.  Results of spin-orbit MRCI calculations are shown.}
\label{Na213_B1}
\begin{tabular}{ccccccc}
 \hline \hline
     $\delta a$  &\hspace{0.2cm} $a$\,(\AA{})& \hspace{0.2cm}  $K$  & \hspace{0.15cm} $J$ &\hspace{0.1cm}  $\Gamma_{xy}$   &  $\Gamma_{zx}$=$-\Gamma_{yz}$ &\hspace{0.1cm}$ |K/J|$\\
      \hline
      \vspace{-0.2cm}\\
     $+2\%$  &\hspace{0.2cm} $3.20$  &\hspace{0.2cm} $\textbf{--16.9}$   & \hspace{0.1cm} $\textbf{4.0}$  &\hspace{0.1cm}  $-0.2$  &  $0.4$& $4.23$ \\
         \vspace{-0.5cm}\\
               \vspace{-0.1cm}\\
     Exp.  &\hspace{0.2cm} $3.14$  & \hspace{0.2cm} $\textbf{--20.8}$   & \hspace{0.1cm} $\textbf{5.2}$  &\hspace{0.1cm}  $-0.7$  &  $0.8$ & $3.98$\\
         \vspace{-0.5cm}\\
           \vspace{-0.1cm}\\
     $-1.5\%$ &\hspace{0.2cm}  $3.09$   & \hspace{0.2cm} $\textbf{--24.6}$   & \hspace{0.1cm} $\textbf{5.9}$  &\hspace{0.1cm}  $-1.3$  &  $1.1$ & $4.13$\\
         \vspace{-0.5cm}\\
               \vspace{-0.1cm}\\
     $-3\%$  &\hspace{0.2cm} $3.04$  & \hspace{0.2cm} $\textbf{--28.9}$   & \hspace{0.1cm} $\textbf{6.8}$  &\hspace{0.1cm}  $-2.3$  &  $1.5$ & $4.27$\\
         \vspace{-0.5cm}\\
           \vspace{-0.1cm}\\
     $-5\%$  &\hspace{0.2cm} $2.98$   & \hspace{0.2cm} $\textbf{--34.7}$   & \hspace{0.1cm} $\textbf{7.7}$  &\hspace{0.1cm}  $-3.4$  &  $2.1$ & $4.50$\\
         \vspace{-0.5cm}\\
         
              \vspace{-0.1cm}\\
        
         \hline
         \hline
      
 \end{tabular}
\vspace{0.3cm}

\caption{NN magnetic couplings (in meV) for bond B2 in Na$_2$IrO$_3$ for variable Ir-Ir bond length $a$, results of spin-orbit MRCI calculations.}
\label{Na213_B2}
\begin{tabular}{ccccccc}
 \hline \hline
     $\delta a$   &\hspace{0.2cm} $a$\,(\AA{})& \hspace{0.2cm}  $K$  &  \hspace{0.15cm} $J$ &\hspace{0.1cm}  $\Gamma_{xy}$   &  $\Gamma_{zx}$=$-\Gamma_{yz}$ &\hspace{0.1cm} $|K/J|$\\
      \hline
      \vspace{-0.2cm}\\
     $+2\%$  & $3.19$  &\hspace{0.2cm} $\textbf{--12.0}$   & \hspace{0.1cm} $\textbf{0.9}$  &\hspace{0.1cm}  $-0.97$  &  $-0.61$& $11.89$ \\
         \vspace{-0.5cm}\\
               \vspace{-0.1cm}\\
     Exp.  & $3.13$  & \hspace{0.2cm} $\textbf{--15.6}$   & \hspace{0.1cm} $\textbf{2.2}$  &\hspace{0.1cm}  $-1.12$  &  $-0.84$ & $7.07$\\
         \vspace{-0.5cm}\\
           \vspace{-0.1cm}\\
     $-1.5\%$  & $3.08$   & \hspace{0.2cm} $\textbf{--18.2}$   & \hspace{0.1cm} $\textbf{3.1}$  &\hspace{0.1cm}  $-1.43$  &  $-0.92$ & $5.89$\\
         \vspace{-0.5cm}\\
               \vspace{-0.1cm}\\
     $-3\%$  & $3.04$  & \hspace{0.2cm} $\textbf{--21.0}$   & \hspace{0.1cm} $\textbf{3.7}$  &\hspace{0.1cm}  $-1.8$  &  $-1.3$ & $5.56$\\
         \vspace{-0.5cm}\\
           \vspace{-0.1cm}\\
     $-5\%$  & $2.97$   & \hspace{0.2cm} $\textbf{--25.6}$   & \hspace{0.1cm} $\textbf{4.8}$  &\hspace{0.1cm}  $-2.5$  &  $-1.7$ & $5.30$\\
         \vspace{-0.5cm}\\
          \vspace{0.1cm}\\

         \hline
         \hline
 \end{tabular}

\end{table}

In both Na$_2$IrO$_3$ and $\alpha$-RuCl$_3$, the NN octahedra display $C_{2h}$ point-group symmetry, which then allows a generalized 
bilinear Hamiltonian of the following form 
 for a pair of pseudospins $i$ and $j$:
\begin{equation}
 {\cal H}^{(\gamma)}_{ij} =J\, \tilde{\bf{S}}_i \cdot \tilde{\bf{S}}_j
           +K \tilde{S}^\gamma_i \tilde{S}^\gamma_j
           +\sum_{\alpha \neq \beta} \Gamma_{\!\alpha\beta}(\tilde{S}^\alpha_i\tilde{S}^\beta_j +
                                                          \tilde{S}^\beta_i \tilde{S}^\alpha_j), \ \
\label{Eq:ham1}
\end{equation} 
where the $\Gamma_{\alpha\beta}$ 
coefficients refer to the off-diagonal components of the symmetric anisotropic exchange matrix, with $\alpha,\beta \in \{x,y,z\}$.
An antisymmetric Dzyaloshinskii-Moriya (DM) exchange is not allowed, given the inversion center for the block of two NN octahedra.

On the other hand, a block of two NN octahedra in the hyperhoneycomb structure may display two different types of point-group symmetry: the so called B2 bonds have $C_{2h}$ point-group symmetry and the Hamiltonian
for these links is given by Eq.\,\eqref{Eq:ham1},
while bond B1 displays $D_2$ point-group symmetry and allows DM antisymmetric anisotropic exchange in the Hamiltonian\,\cite{Katukuri16}.
However, only the $x$-component of the DM vector is non-zero. The Hamiltonian for bond B1 can be then written as: 
\begin{equation}
 { \cal {\bar H}}^{(z)}_{ij} =J\, \tilde{\bf{S}}_i \cdot \tilde{\bf{S}}_j
           +K \tilde{S}^z_i \tilde{S}^z_j
           + \Gamma_{xy}(\tilde{S}^x_i\tilde{S}^y_j +
                                                         \tilde{S}^y_i \tilde{S}^x_j) + 
                                                                              {\vec D} \cdot {\tilde {\bf S}}_i\times {\tilde {\bf S}}_j.
\label{Eq:ham2}
\end{equation}
 A local Kitaev reference frame is used here, such that for each TM-TM link, the $z$-coordinate is perpendicular to the TM$_2$L$_2$ plaquette.
 Mapping of the {\it ab initio} data onto an effective spin Hamiltonian is carried out following the procedure earlier used in Refs.\,\cite{Yadav16,Bogdanov15,Yadav17}.
Experimental crystallographic data were used in the present calculations for Na$_2$IrO$_3$, $\beta$-Li$_2$IrO$_3$, and $\alpha$-RuCl$_3$ as reported in \cite{Choi12,Takayama15,nagler16}, respectively.
To test the qualitative trends found phenomenologically for the rescaling of the coupling constants, we further considered  
structural data corresponding to $-1.5\%$, $-3\%$, $-5\%$, and $+2\%$ change in the TM-TM bond length in the many-body quantum-chemistry calculations. 
Further details of the calculations are provided in SM.

\begin{table}[t]
\caption{NN magnetic couplings (in meV) in RuCl$_3$ for variable Ru-Ru bond length $a$, results of spin-orbit MRCI calculations.}
\begin{center}
\label{RuCl3}
\begin{tabular}{cccccccc}
 \hline \hline
     $\delta a$  & \hspace{0.2cm} $a$\,(\AA{})& \hspace{0.2cm}  $K$  &\hspace{0.1cm} $J$ &\hspace{0.1cm}  $\Gamma_{xy}$   &\hspace{0.1cm}  $\Gamma_{zx} = -\Gamma_{yz}$ &\hspace{0.1cm} $ |K/J|$ \\
      \hline
      \vspace{-0.2cm}\\
     $+2\%$  & $3.52$  &\hspace{0.2cm} $\textbf{--4.1}$   &\hspace{0.1cm} $\textbf{0.8}$  &\hspace{0.1cm}  $-0.9$  &  $-0.4$ & $5.31$\\
         \vspace{-0.5cm}\\
               \vspace{-0.1cm}\\
     Exp.  & $3.45$  & \hspace{0.2cm} $\textbf{--5.6}$   &\hspace{0.1cm} $\textbf{1.2}$  &\hspace{0.1cm}  $-1.2$  &  $-0.7$ & $4.67$ \\
         \vspace{-0.5cm}\\
           \vspace{-0.1cm}\\
     $-1.5\%$  & $3.40$   & \hspace{0.2cm} $\textbf{--7.1}$   &\hspace{0.1cm} $\textbf{1.8}$  & \hspace{0.1cm} $-1.3$  &  $-0.9$ & $3.99$\\
         \vspace{-0.5cm}\\
               \vspace{-0.1cm}\\
     $-3\%$  & $3.35$  & \hspace{0.2cm} $\textbf{--8.7}$   &\hspace{0.1cm} $\textbf{2.3}$  &\hspace{0.1cm}  $-1.6$  &  $-1.2$ & $3.78$\\
         \vspace{-0.5cm}\\
           \vspace{-0.1cm}\\
     $-5\%$  & $3.28$   & \hspace{0.2cm} $\textbf{--11.4}$   &\hspace{0.1cm} $\textbf{2.8}$  &\hspace{0.1cm}  $-2.0$  &  $-1.8$ & $4.05$\\
         \vspace{-0.5cm}\\
         
              \vspace{-0.1cm}\\
         
         \hline
         \hline
      
 \end{tabular}
\end{center}
\end{table}

{\it Results.}
 We start our discussion on the variations of the magnetic exchange interactions when modifying bond lengths with the case of Na$_2$IrO$_3$.
NN magnetic couplings as derived from spin-orbit multireference configuration-interaction (MRCI) calculations~\cite{Helgaker2000} are listed in Table \ref{Na213_B1}. 
For bond B1, $K$ increases from $-20.8$ meV for to the experimental crystal structure at ambient pressure to $-35.7$ meV on 5\% reduction 
of the Ir-Ir bond length. $J$, on the other hand, displays a rather modest enhancement, from $5.2$ to $7.7$ meV. This translates in an increase of the $|K/J|$ ratio from 3.98 to 4.50. 
$\Gamma_{xy}$ and $\Gamma_{zx}$ also gain significant strength with rising pressure but remain nevertheless one order of magnitude smaller than $K$.

In the case of bond B2, the trends look a bit different: while $K$ evolves in a similar fashion as for bond B1, $J$ becomes almost twice the value at ambient pressure for the shortest Ir-Ir bond length considered here. 
As a consequence, $|K/J|$ decreases for bond B2 (see Table \ref{Na213_B2}).
However, the $|K/J|$ ratio jumps from 7 at ambient pressure to 12 for 2\% elongation of the Ir-Ir bond.
Such increased bond lengths could be realized under tensile strain.
The steep rise of the $|K/J|$ ratio can be understood as a result of the rapid downturn of the Heisenberg $J$ towards 0.
In fact, such a decreasing trend in $J$ suggests that it would completely vanish with further slight elongation of the bonds,
which then would lead to a Hamiltonian of pure anisotropic nature. 
The fact that the two distinct links in Na$_2$IrO$_3$ show different relative gain in $K$ and $J$ with the change in Ir-Ir distance indicates
that the strength of the spin-spin couplings and various exchange processes for each bond are additionally
significantly controlled by other attributes of the local environment such as the Ir-O-Ir angle.  

\begin{table}[b] 
\caption{NN magnetic couplings (in meV) for bond B1 in $\beta$-Li$_2$IrO$_3$ for variable Ir-Ir bond length $a$, results of spin-orbit MRCI calculations.}
\label{Li213_B1}
\begin{center}
\vspace{-0.2cm}
\begin{tabular}{ccccccc}
 \hline \hline
     $\delta a$  &\hspace{0.2cm} $a$\,(\AA{}) & \hspace{0.3cm}  $K$  &\hspace{0.2cm} $J$ &\hspace{0.2cm} $D_x$   &\hspace{0.2cm}  $\Gamma_{xy}$  \\
      \hline
      \vspace{-0.2cm}\\
     $+2\%$  &\hspace{0.2cm} $3.04$ &\hspace{0.3cm} $\textbf{--11.70}$   &\hspace{0.2cm} $\textbf{0.21}$  &\hspace{0.2cm}  $0.30$  &\hspace{0.2cm}  $-1.69$  \\
         \vspace{-0.5cm}\\
               \vspace{-0.1cm}\\
     Exp.  &\hspace{0.2cm} $2.98$  & \hspace{0.3cm} $\textbf{--14.78}$   &\hspace{0.2cm} $\textbf{-0.26}$  &\hspace{0.2cm}  $0.35$  &\hspace{0.2cm}  $-2.08$ \\
         \vspace{-0.5cm}\\
           \vspace{-0.1cm}\\
       $-3\%$  &\hspace{0.2cm} $2.89$  & \hspace{0.3cm} $\textbf{--17.01}$   &\hspace{0.2cm} $\textbf{-0.41}$  & \hspace{0.2cm} $0.45$  &\hspace{0.2cm}  $-3.48$ \\
         \vspace{-0.5cm}\\
           \vspace{-0.1cm}\\
     $-5\%$  &\hspace{0.2cm} $2.83$    & \hspace{0.3cm} $\textbf{--20.72}$   &\hspace{0.2cm} $\textbf{-0.60}$  &\hspace{0.2cm}  $0.56$  & \hspace{0.2cm} $-4.80$ \\
         \vspace{-0.5cm}\\
         
              \vspace{-0.1cm}\\
         
         \hline
         \hline
      
 \end{tabular}
\end{center}
\vspace{-0.2cm}
\caption{NN magnetic couplings (in meV) for bond B2 in $\beta$-Li$_2$IrO$_3$ for variable Ir-Ir bond length $a$, results of spin-orbit MRCI calculations.}
\label{Li213_B2}
\begin{center}
\vspace{-0.3cm}
\begin{tabular}{ccccccc}
 \hline \hline
   $\delta a$  &\hspace{0.2cm} $a$ (\AA{})& \hspace{0.2cm}  $\textbf{K}$  &\hspace{0.2cm} $\textbf{J}$ &\hspace{0.2cm}  $\Gamma_{xy}$   &  $\Gamma_{zx} = -\Gamma_{yz}$ &\hspace{0.2cm}$ |K/J|$ \\
      \hline
      \vspace{-0.2cm}\\
     $+2\%$  & $3.03$  &\hspace{0.2cm} $\textbf{--11.1}$   &\hspace{0.2cm} $\textbf{-0.9}$  &\hspace{0.2cm}  $-3.0$  &  $-0.7$ & $11.85$\\
         \vspace{-0.5cm}\\
               \vspace{-0.1cm}\\
     Exp.  & $2.97$  & \hspace{0.2cm} $\textbf{--12.2}$   &\hspace{0.2cm} $\textbf{-2.1}$  &\hspace{0.2cm}  $-4.1$  &  $-1.0$ & $5.81$\\
         \vspace{-0.5cm}\\
           \vspace{-0.1cm}\\
     $-1.5\%$  & $2.93$   & \hspace{0.2cm} $\textbf{--14.1}$   &\hspace{0.2cm} $\textbf{-2.7}$  &\hspace{0.2cm}  $-4.9$  &  $-1.1$& $5.20$ \\
         \vspace{-0.5cm}\\
               \vspace{-0.1cm}\\
     $-3\%$  & $2.88$  & \hspace{0.2cm} $\textbf{--15.6}$   &\hspace{0.2cm} $\textbf{-3.2}$  & \hspace{0.2cm} $-6.1$  &  $-1.3$& $4.88$ \\
         \vspace{-0.5cm}\\
           \vspace{-0.1cm}\\
     $-5\%$  & $2.82$   & \hspace{0.2cm} $\textbf{--17.7}$   & \hspace{0.2cm} $\textbf{-3.8}$  &\hspace{0.2cm}  $-8.1$  &  $-1.7$& $4.57$ \\
         \vspace{-0.5cm}\\
         
              \vspace{-0.1cm}\\
         
         \hline
         \hline
      
 \end{tabular}
\end{center}
\end{table}

\begin{figure}
\label{fit}
\begin{minipage}{0.5\textwidth}
\hspace{-0.7cm}
 \includegraphics[angle=270, width=7.2cm]{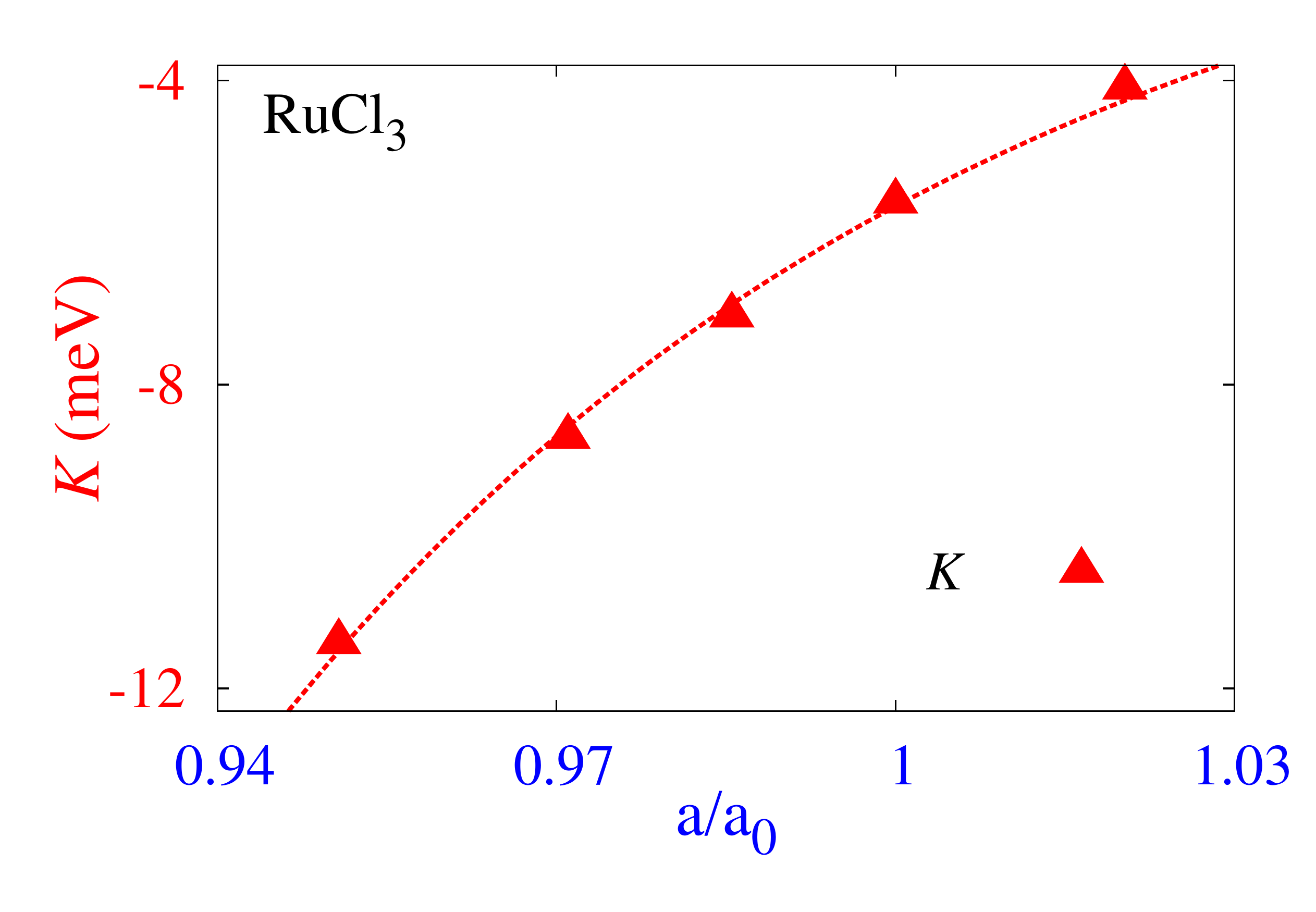}
\end{minipage}%
\vspace{-0.2cm}
\begin{minipage}{0.5\textwidth}
\hspace{-0.7cm}
 \includegraphics[angle=270, width=7.2cm]{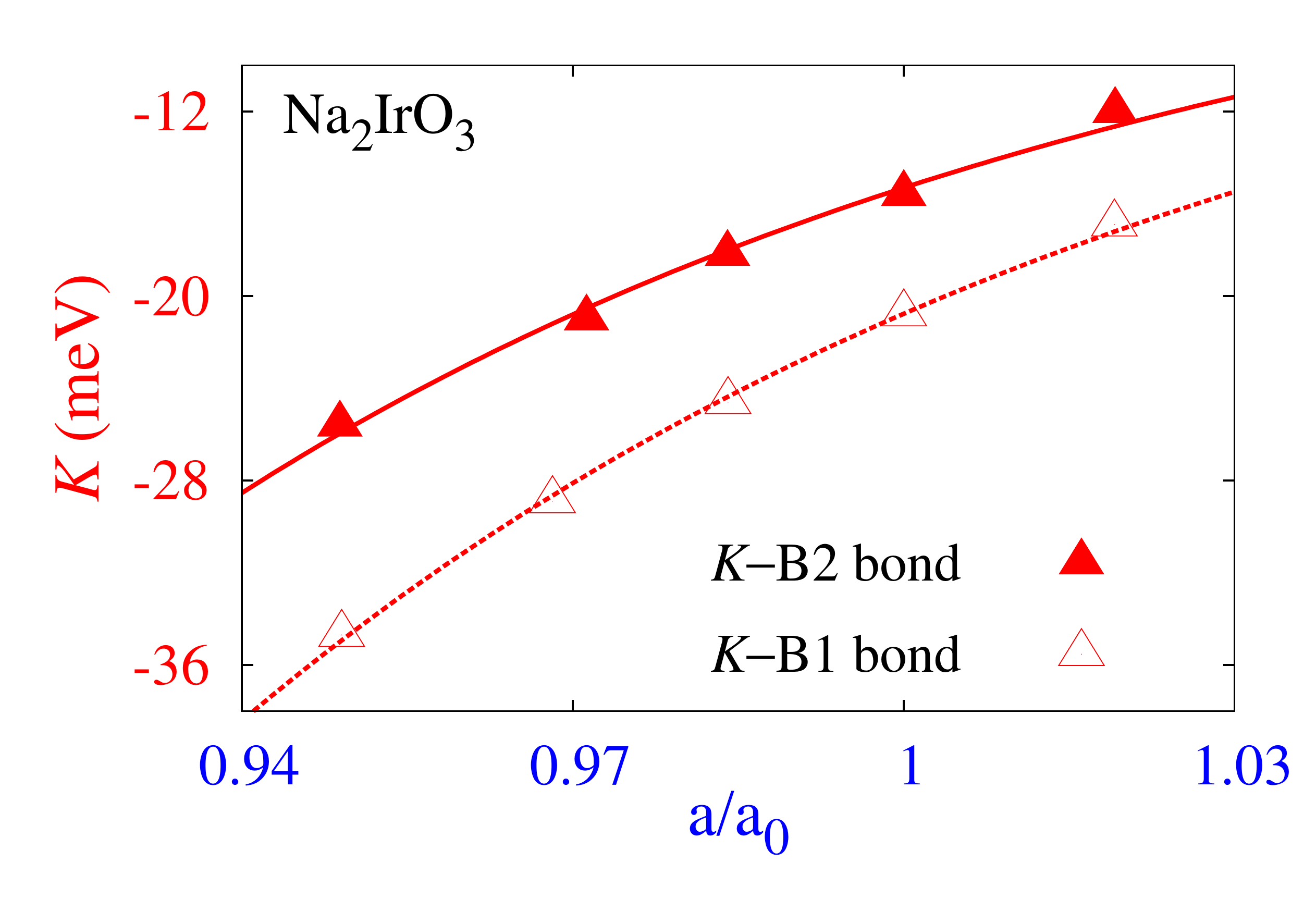}
\end{minipage}
\vspace{-0.2cm}
\caption{NN Kitaev coupling for variable TM-TM bond length, fitted with the function $K=K_0\,x^n$; plots for $\alpha$-RuCl$_3$ (top) and for B1 and B2 links in Na$_2$IrO$_3$ (bottom) are shown.}
\label{fig:K-dependence}
\end{figure}

The variations of the NN magnetic exchange interactions with modifying bond lengths in $\alpha$-RuCl$_3$, as obtained by spin-orbit MRCI calculations,
are listed in Table \ref{RuCl3}. $K$ remains on the ferromagnetic side and increases to $-11.5$ meV on 5\% reduction of the Ru-Ru distance as compared
to the value of $-5.6$ meV at ambient pressure. For the same case, $J$ moves to 2.8 meV from a value of 1.2 meV at normal pressure.
An interesting point to note is again the reduction of $J$ 
towards zero with 2\% elongation of the Ru-Ru bond. For stretched bonds, the $|K/J|$ ratio reaches in fact the largest value. 
$\Gamma_{xy}$ and $\Gamma_{zx}$ also display a strong dependence on interatomic distances but these effective parameters are never larger than
25\% of $K$ in RuCl$_3$. 
In contrast, in
$\beta$-Li$_2$IrO$_3$, $\Gamma_{xy}$ becomes as large as half the value of $K$ and twice the value of $J$ for the shortest Ir-Ir distance considered
for bond B2 (see Tables \ref{Li213_B1} and \ref{Li213_B2}).
This large $\Gamma_{xy}$ stands out while comparing trends with other honeycomb systems.

All NN magnetic couplings computed for $\beta$-Li$_2$IrO$_3$ are listed in Tables \ref{Li213_B1} and \ref{Li213_B2}.
For the case of B2 links in $\beta$-Li$_2$IrO$_3$, $K$ rises to $-17.7$ meV on 5\% cutback in the Ir-Ir distance, an increase by nearly 50\%
as compared to $-12.2$ meV at
ambient pressure. $J$, on the other hand, changes from $-2.1$ meV at ambient conditions to a value of
$-3.8$ meV.
Similar to the other compounds, the $|K/J|$ ratio is maximum for positive bond-length increments. 
$J$ even changes sign for 2\% increase of the Ir-Ir distance for bond B1, which suggests that applying a very modest amount of
tensile strain might bring the system close to the $J=0$ limit, where only the anisotropic couplings are finite.

To determine the order of the exponential dependence of $K$ as a function of changes in the TM-TM distance, the trends shown in the tables were fitted to the function $K=K_0\,x^n$, where
$K_0$ represents the Kitaev exchange amplitude at ambient pressure and $n$ refers to the exponent of fractional change in the TM-TM distance, i.e., $a/a_0 = 1 + \delta a$.
The plots shown in Fig.\,\ref{fig:K-dependence} display the variation of $K$ in Na$_2$IrO$_3$ and $\alpha$-RuCl$_3$ fitted to such a function.
Using these fits, the value of the exponent is determined to be $n=-15$ 
 for the case of RuCl$_3$.
The exponent decreases to the values of $n=-13$ and $-11$ for the cases of the bonds B2 and B1 in Na$_2$IrO$_3$, respectively, while
it becomes as small as $n=-8$ for the Ir-Ir bonds in $\beta$-Li$_2$IrO$_3$.
The value of the exponent for $\alpha$-RuCl$_3$ fits very nicely to the values predicted by the simplistic picture mentioned in the 
qualitative analysis section. The lower values of exponent obtained in the cases of Na$_2$IrO$_3$ and $\beta$-Li$_2$IrO$_3$
points to the changing nature of the exchange processes with slight modification of the surroundings, such as having different bond angles and ligand charge.


{\it Summary.\, } We have employed advanced quantum-chemistry methods to model the effects of uniform pressure and strain  on the exchange couplings in iridium and ruthenium compounds with honeycomb and related lattices.
The obtained results demonstrate that the Kitaev, Heisenberg, symmetric off-diagonal, and antisymmetric anisotropic magnetic interactions stemming from the different exchange processes   renormalize differently under volume change. This, in turn, suggests that 
by introducing external pressure or strain on actual materials one could experimentally explore the extremely rich theoretical phase diagram composed of the quantum-spin liquid, collinear, as well as non-collinear ordered states. We believe that the present results  are relevant to the  pressure-induced melting of the magnetic long-range order experimentally suggested in $\beta$-Li$_2$IrO$_3$ and $\alpha$-RuCl$_3$ compounds ~\cite{Takayama15,Veiga17,Wang17} and will motivate further pressure and strain experiments on Kitaev materials.

This work is supported by the DFG through SFB 1143.
GJ, SR, and JvdB benefitted from the facilities of the KITP. 
GJ was supported in part by the NSF under Grant No.\ NSF PHY11-25915.
RY and LH acknowledge Ulrike Nitzsche for technical support.

\bibliography{Pressure_dep}

\end{document}